\documentclass[aps,twocolumn,epsf,showpacs]{revtex4} % PH. REV.
\usepackage{graphicx}
\newcommand{\beq}{\begin{equation}}
\newcommand{\eeq}{\end{equation}}
\newcommand{\ba}{\begin{array}}
\newcommand{\ea}{\end{array}}
 
\begin{document}

\title{Mixing-demixing transition and collapse of a vortex state \\
in a quasi-two-dimensional boson-fermion mixture}
 
\author{Sadhan K. Adhikari$^1$\footnote{E-mail: 
adhikari@ift.unesp.br; URL: http://www.ift.unesp.br/users/adhikari} 
and Luca Salasnich$^2$\footnote{E-mail: salasnich@pd.infn.it; 
URL: http://www.padova.infm.it/salasnich}}
\affiliation
{$^1$Instituto de F\'{\i}sica Te\'orica, UNESP - S\~ao Paulo State 
University, 01.405-900 S\~ao Paulo, S\~ao Paulo, Brazil
\\  
$^2$CNISM and CNR-INFM, Unit\`a di Padova, 
Dipartimento di Fisica ``Galileo Galilei'', Universit\`a di Padova, 
Via Marzolo 8, 35131 Padova, Italy} 
 
\date{\today}
 
\begin{abstract} We investigate the mixing-demixing transition and the
collapse in a quasi-two-dimensional degenerate boson-fermion mixture
(DBFM) with a bosonic vortex.  We solve numerically a
quantum-hydrodynamic model based on a new density functional which
accurately takes into account the dimensional crossover.  It is
demonstrated that with the increase of interspecies repulsion, a mixed
state of DBFM could turn into a demixed state.  The system collapses for
interspecies attraction above a critical value which depends on the
vortex quantum number.  For interspecies attraction just below this
critical limit there is almost complete mixing of boson and fermion
components. Such mixed and demixed states of a DBFM could be
experimentally realized by varying an external magnetic field near a
boson-fermion Feshbach resonance, which will result in a continuous
variation of interspecies interaction. \end{abstract}

\pacs{03.75.Lm, 03.75.Ss}

\maketitle
 
\section{Introduction}

A quantum degenerate Fermi gas (DFG) cannot be achieved 
by evaporative cooling due to a strong repulsive Pauli-blocking 
interaction at low energies among spin-polarized fermions \cite{exp1}. 
Trapped DFG has been achieved only by sympathetic cooling 
in the presence of a second boson or fermion component. 
Recently, there have been successful observation 
\cite{exp1,exp2,exp3,exp4} and associated experimental 
\cite{exp5,exp5x,exp6} and theoretical 
\cite{yyy,yyy1,yyy2,md1,md2,zzz,capu,anna-minguzzi,ska}
studies of degenerate boson-fermion mixtures by different experimental
groups \cite{exp1,exp2,exp3,exp4} in the following systems: 
$^{7}$Li-$^{6}$Li \cite{exp3}, 
$^{23}$Na-$^6$Li \cite{exp4} and $^{87}$Rb-$^{40}$K 
\cite{exp5,exp5x}. Moreover, there have been studies 
of a degenerate mixture of two components of fermionic 
$^{40}$K \cite{exp1} and $^6$Li \cite{exp2} 
atoms. The collapse of the DFG in a degenerate boson-fermion mixture 
(DBFM) of $^{87}$Rb-$^{40}$K has been observed and studied 
by Modugno {\it et al.} \cite{exp5,zzz,ska}, and has also been 
predicted in a degenerate fermion-fermion mixture of different-mass 
atoms \cite{skac}. 
 
Several theoretical investigations \cite{md1,md2,yyy1} of a trapped DBFM
considered the phenomenon of mixing-demixing in a state of zero angular
momentum when the boson-fermion
repulsion is increased. For a weak boson-fermion repulsion both 
a Bose-Einstein condensate (BEC) and a DFG 
have maxima of probability density at the center of the harmonic trap.
However, with the increase of boson-fermion repulsion, the maximum of
the probability density of the DFG could be slowly expelled from the
central region. With further increase of boson-fermion repulsion, the
DFG could be completely expelled from the central region which will
house only the BEC. This phenomenon has been termed mixing-demixing in a
DBFM. The phenomenon of demixing has drawn some attention lately as in
a demixed state an exotic configuration of the mixture is formed, where
there is practically no overlap between the two components and one can
be observed and studied independent of the other. It has been argued
\cite{yyy1} that such a demixed state in a DBFM should be possible
experimentally by increasing the interspecies scattering length near a
Feshbach resonance \cite{fesh}. More recently mixing-demixing has been 
studied in a degenerate fermion-fermion mixture \cite{mdmff}.

On the other hand if the interspecies interaction is
turned attractive and its strength increased, the DBFM  collapses above 
a critical strength. There has been several experimental 
\cite{exp5,bongs,ccol} and theoretical 
\cite{ska,zzz,yyy} studies of collapse in a DBFM 
of $^{40}$K-$^{87}$Rb mixture. As the interaction 
in a pure DFG at short distances is repulsive due to Pauli
blocking, there cannot be a collapse in it. A collapse is possible in a 
DBFM in the presence of a sufficiently strong Boson-fermion attraction 
which can overcome the Pauli repulsion among identical fermions 
\cite{exp5}. 

It is pertinent to see how mixing-demixing manifest for 
interspecies attraction below the critical value for collapse. 
The appearance 
of a quantized bosonic vortex state is the genuine confirmation of
superfluidity in a trapped BEC. 
In view of many experimental studies of such bosonic vortex states
it is also of interest to see how the mixing-demixing 
phenomenon in a DBFM modify in the presence of a bosonic vortex. 
The vortices are quantized rotational excitations \cite{donn} and can be 
observed in two-dimensional (2D) systems. The lowest of such excitations 
with unit angular momentum ($\hbar$) per atom is the nonlinear 
extension of a well-understood linear quantum state \cite{skal}. 
Vortex states in a BEC have been observed experimentally 
\cite{exptv}. Different techniques for 
creating vortex states in BEC have been suggested \cite{d1}, e.g.,
stirring the BEC with an external laser \cite{d2}, forming spontaneously
in evaporative cooling \cite{d3}, using a ``phase imprinting method"
\cite{d4}, and rotating an axially symmetric trap \cite{d5}. 
Recently, the stability of the vortex state and the formation of persistent 
currents have been theoretically analyzed also in toroidal traps 
\cite{rokhsar,sala-toro,dalfovo-toro}.
%new sentence added 
The generation of vortex in degenerate fermions is much more complex 
\cite{karx}
and 
we shall 
not consider this possibility here.

The purpose of this paper is to study and illustrate the mixing-demixing
phenomenon for both attractive and repulsive interspecies interaction in a
trapped DBFM vortex in a quasi-2D configuration using a
quantum-hydrodynamic model inspired by the 
success of this model in the investigation of fermionic collapse 
\cite{ska} and bright \cite{fbs2,sala-solitonBF} 
and dark \cite{fds} solitons in a DBFM. 
The conclusions of the study on bright soliton \cite{fbs2} are in 
agreement with a microscopic study \cite{bongs1}, and those on collapse
\cite{ska} are in agreement with experiments \cite{exp5,bongs}. This
time-dependent mean-field-hydrodynamic model was suggested recently
\cite{ska} to study the collapse dynamics of a DBFM.

In addition to the study of mixing-demixing in a DBFM in a 
quasi-2D configuration, we also study conditions of 
stability and collapse in it. Specifically, we study the conditions of 
stability when the parameters of a DBFM are modified, e.g., boson-boson 
and boson-fermion interactions as well as the boson and fermion numbers.

There have been prior investigations of 
mixing-demixing \cite{md1,md2} in a trapped DBFM 
upon an increase of interspecies repulsion. Also, there have been 
previous investigations of stability and collapse 
\cite{zzz,anna-minguzzi} in a trapped DBFM.
In contrast to these previous time-independent studies for stationary
states, the present study relies on a time-dependent formulation and
investigates mixing-demixing and stability and collapse for both 
attractive and repulsive interspecies interaction and extends to the case 
of vortex states for the first time.

The model hydrodynamic equations  in a quasi-2D form is 
derived from a Lagrangian density for the DBFM where the boson 
Lagrangian is taken in the usual mean-field Gross-Pitaevskii form 
\cite{stringari-book}. The interaction Lagrangian between bosons and 
fermions is 
also taken to have the standard product form of boson and fermion 
probability densities. We derive a new Lagrangian for the quasi-2D 
fermions by putting them in a box of length $L$ along $x$ and $y$ 
directions and in a harmonic potential well  along $z$ direction. By 
occupying 
the lowest single-particle fermion states we calculate the fermion 
probability density as a function of chemical potential from which we 
obtain the corresponding fermion Lagrangian density. The resultant 
hydrodynamic equations  have a nonpolynomial nonlinearity for the 
fermions, which we use in our calculation. In the strict 2D limit, when 
axial excitations are not 
allowed, this nonlinearity reduced to a standard cubic  
form.

The paper is organized as follows. 
In Sec. II we present an account of the quantum hydrodynamic model 
consisting of a set of coupled partial differential equations involving 
a quasi-2D BEC and a DFG. In Sec. III we present the numerical results 
on mixing-demixing and collapse of a DBFM in two subsections, 
respectively. In Sec. IV we present a summary and discussion. 
Some technical details are given in Appendix.  
 
\section{Boson-fermion Lagrangian for quasi-2D hydrodynamics} 

We consider a DBFM with $N_B$ Bose-condensed atoms 
of mass $m_B$ and $N_F$ spin-polarized 
fermions of mass $m_F$ at zero temperature. 
A natural choice for a quasi-2D trap-geometry is a very 
strong confinement along the $z$ axis: in this axial direction 
we choose a harmonic potential of frequency $\omega_z$. 
In the cylindric radial directions 
we take two generic external potentials for 
bosons and fermions: $V_B(\rho)$ and $V_F(\rho)$, 
where $\rho=(x^2+y^2)^{1/2}$ is the cylindric radial coordinate. 
 
To describe this quasi-2D DBFM we use two dynamical fields:  
$\psi_B(\rho,t)$ and $\psi_F(\rho,t)$. The complex function 
$\psi_B(\rho,t)$ is the hydrodynamic field of the Bose gas, 
such that $n_B=|\psi_B|^2$ is the 2D bosonic probability density 
and $v_B=i\partial_{\rho} \ln{\left( \psi_B/|\psi_B| \right)}$ 
is the bosonic velocity. The complex function
$\psi_F(\rho,t)$ is the hydrodynamic field of the Fermi gas,
such that $n_F=|\psi_F|^2$ is the 2D fermionic density 
and $v_F=i\partial_{\rho}
\ln{\left( \psi_F/|\psi_F| \right)}$ 
is the fermionic velocity. These two complex fields are the Lagrangian 
variables of the Lagrangian density 
\beq 
{\cal L} = {\cal L}_B + {\cal L}_F + {\cal L}_{BF} \; , 
\label{yy} 
\eeq
where ${\cal L}_B$ is the bosonic Lagrangian, 
${\cal L}_F$ is the fermionic Lagrangian and 
${\cal L}_{BF}$ is the Lagrangian of the boson-fermion interaction. 
It is important to stress that in our model, 
based on quantum hydrodynamics 
\cite{stringari-book, lipparini}, the bosonic Lagrangian 
${\cal L}_B$ describes very accurately all 
the dynamical properties of the dilute quasi-2D BEC 
\cite{lipparini,sala-npse}, while 
the fermionic Lagrangian ${\cal L}_F$ can be safely used 
only for static and collective properties of the 
quasi-2D Fermi gas \cite{lipparini}. Note that recently quantum 
hydrodynamics has been also successfully applied to investigate 
the dimensional crossover from a 3D BEC 
to a 1D Tonks-Girardeau gas \cite{sala-3DBEC-1DTG,11} in  a DBFM. 

The bosonic Lagrangian is given by \cite{ska,fbs2,sala-npse} 
\begin{eqnarray} 
{\cal L}_B &=& \frac{i\hbar}{2}(\psi_B^* \partial_t
\psi_B- \psi_B \partial_t
\psi_B^*)
-\frac{\hbar^2}{2m_B}|\nabla_\rho \psi_B|^2
\nonumber \\ 
&-& 
{\hbar^2 l^2 \over 2m_B \rho^2}  n_B 
- {\cal E}_B( n_B ) - V_Bn_B \,  ,
\label{lagrange-bose}
\end{eqnarray}
where $\hbar^2 l^2/(2m_B \rho^2)$ is the centrifugal term 
of the bosonic vortex, $l$ is the integer quantum number of circulation 
and $\hbar l$ is the angular momentum of each atom in the axial ($z$) 
direction \cite{skal}. 
The term ${\cal E}_B(n_B)$ is the bulk energy 
density of the dilute and interacting 
quasi-2D BEC under axial 
harmonic confinement. As shown in Ref. \cite{sala-npse}, this bulk energy 
density is a nonpolynomial function of the 2D bosonic 
density $n_B=|\psi_B|^2$. 
For small bosonic densities, i.e. for 
$0\leq n_B < 1/(2\sqrt{2\pi} a_{BB} a_{zB})$ where $a_{BB}$ is 
the 3D Bose-Bose scattering length and 
$a_{zB}=\sqrt{\hbar/(m_B \omega_z )}$ is the 
characteristic length
of axial harmonic confinement for bosons, the BEC is strictly 2D and one 
finds 
\beq \label{z1}
{\cal E}_B = {1\over 2} g_{BB}\, n_B^2 \, , 
\eeq 
where $g_{BB}=4\pi\hbar^2a_{BB}/(\sqrt{2\pi} a_{zB} m_B)$ 
is the 2D inter-atomic strength \cite{sala-npse}. 
For very large densities, i.e. for 
$n_B \gg 1/(2\sqrt{2\pi}a_{BB} a_{zB})$, the BEC is instead 3D 
and ${\cal E}_B$ scales as $n_B^{5/3}$ 
(for details see Ref. \cite{sala-npse}, where 
nonpolynomial Schr\"odinger equations are derived 
for cigar-shaped and disk-shaped BECs starting 
from the 3D Gross-Pitaevskii Lagrangian). 
Here we consider a BEC with a small $a_{zB}$ (strong axial 
confinement) and a much smaller scattering length $a_{BB}$  
and so the BEC is strictly 2D. 

The fermionic Lagrangian is given by \cite{ska,fbs2}
\begin{eqnarray} 
{\cal L}_F &=& \frac{i\hbar}{2}(\psi_F^* \partial_t
\psi_F- \psi_F \partial_t\psi_F^*)
-\frac{\hbar^2}{6m_F}|\nabla_\rho \psi_F|^2
\nonumber \\ 
&-& 
{\cal E}_F( n_F ) - V_Fn_F \,  ,
\label{lagrange-fermi}
\end{eqnarray} 
where ${\cal E}_F(n_F)$ is the bulk energy
density of a non-interacting quasi-2D Fermi gas 
at zero temperature and under axial harmonic confinement. 
As shown in Appendix, 
this bulk energy density is a nonpolynomial function of 
the 2D fermionic density $n_F=|\psi_F|^2$. 
For small fermionic densities, 
i.e. for $0\le n_F < 1/(2\pi a_{zF}^2)$, 
the Fermi gas is strictly 2D and one finds 
${\cal E}_F = \hbar\omega_z \pi (a_{zF} n_F)^2$, where 
$a_{zF} =\sqrt{\hbar/( \omega_z m_F)}$ 
is the characteristic length of axial 
harmonic confinement of fermions. 
For large densities, i.e. for $n_F \gg 1/(2\pi a_{zF}^2)$, 
the Fermi gas is 3D and, as shown in Appendix, 
one has ${\cal E}_F = (4a_{zF}^3\sqrt{\pi}/3)(\hbar \omega_z) 
n_F^{3/2}$. Contrary to the case of bosons, whose 
dimensionality depends also on $a_{BB}$ (that is very small), 
for fermions it is necessary to use an extremely small 
$a_{zF}$ and a very small number of atoms to 
have a strictly 2D configuration. 
This is not the case of real experiments and so we use the formula 
\beq
{\cal E}_F  = {\hbar \omega_z\over a_{zF}^2} 
\left\{ \ba{ll} 
\pi (n_F a_{zF}^2)^2 \quad  
\mbox{for} \quad 0\le n_F a_{zF}^2 < {1\over 2\pi} \\
{1\over 6\pi}(4\pi n_F a_{zF}^2 - 1)^{3/2} + {1\over 12 \pi}
\quad \mbox{for} \quad n_F a_{zF}^2 \ge {1\over 2\pi} \, , 
\ea \right.
\eeq
which has been deduced in Appendix and gives the full 2D-3D crossover 
of an ideal Fermi gas that is uniform in the cylindric radial direction 
and under harmonic confinement in the cylindric axial direction. 
It is interesting to stress that the study of 2D-3D (and 
1D-3D)
cross overs have a long history in trapped (bosonic) atoms. 
There have been careful studies of a pair of 
of trapped atoms \cite{ref} as well as of a large number of trapped 
atoms \cite{ref1}. In the Appendix we consider a different type of the 
2D-3D
crossover for a large number of ideal Fermi gas atoms  distributed 
over different quantum states obeying Pauli principle.

In the fermionic Lagrangian of Eq. (\ref{lagrange-fermi}) 
the Weisz\"acker gradient term 
$-\hbar^2|\nabla_{\rho}\psi_F|^2 /(6m_F)$ 
takes into account the additional kinetic energy 
due to spatial variation \cite{capu} 
but contributes little to this problem compared to the dominating 
Pauli-blocking term ${\cal E}_F(n_F)$ \cite{jz,pi} for a large number 
of Fermi atoms. 
The interaction between intra-species fermions in the 
spin-polarized state is highly suppressed due
to the Pauli-blocking term and has been neglected
in the Lagrangian ${\cal L}_F$ and will be neglected throughout.

Finally, the Lagrangian of the boson-fermion interaction reads 
\cite{capu,sala-boris}
\beq 
{\cal L}_{BF} = - g_{BF}\, n_B \, n_F \; , 
\label{lagrange-bose-fermi} 
\eeq
where $g_{BF}=2\pi \hbar^2 a_{BF}/(m_R 
\sqrt{2\pi\, a_{zB}\, a_{zF}} )$ with $a_{BF}$ 
the 3D Bose-Fermi scattering length and 
$m_R=m_Bm_F/(m_B+m_F)$ the Bose-Fermi reduced mass. 

The Euler-Lagrange equations of motion of 
the Lagrangian density (\ref{yy}) with Eqs. 
(\ref{lagrange-bose}), (\ref{lagrange-fermi}),
and (\ref{lagrange-bose-fermi}) 
are given by 
\beq 
\label{e} 
i\hbar\frac{\partial }{\partial t} \psi_B = 
\biggr[ -\frac{\hbar^2\nabla_{\rho}^2}{2 m_B} 
+ {\hbar^2l^2\over 2m_B \rho^2 } + \mu_B(n_B) + 
V_B + g_{BF} n_F \biggr] \psi_B \; ,
\eeq
\beq
\label{f}
i\hbar\frac{\partial}{\partial t} \psi_F = 
\biggr[-\frac{\hbar^2\nabla_{\rho}^2}{6m_F} + 
\mu_F(n_F) + V_F + g_{BF} n_B 
\biggr]\psi_F \; ,
\eeq
where $\mu_B = {\partial {\cal E}_B/ \partial n_B} = 
g_{BB}n_B$ is the bulk 
chemical potential of the strictly 2D BEC and 
$\mu_F = {\partial {\cal E}_F/ \partial n_F}$ is the bulk chemical 
potential, given by Eq. (\ref{chem-fermi}) of Appendix, 
of the ideal Fermi gas in the 2D-3D crossover. 
The normalization used in Eqs. (\ref{e}) and (\ref{f})
and above is $2 \pi \int_0^\infty |\psi_j|^2 \rho d\rho = N_j$. 

For $g_{BF}=0$  Eq. (\ref{e}) is the 
2D Gross-Pitaevskii equation while Eq. (\ref{f}) is 
essentially a time-dependent generalization of a  quasi-2D version of 
the time-independent 
equations of motions 
suggested by Capuzzi {\it at al} \cite{capu} and 
Minguzzi {\it et al.} \cite{anna-minguzzi} 
to study static and collective properties of a confined, 
dilute and spin-polarized Fermi gas. 
That time-independent version was a generalization of the Thomas-Fermi 
(TF)
approximation for the density of a Fermi gas \cite{jz}. The   
TF 
approximation 
for a stationary Fermi gas can be obtained from Eq. (\ref{f}) by setting 
the 
kinetic energy term to zero. For a large number of Fermi atoms, in Eq. 
(\ref{f}) the  nonlinear term $\mu_F(n_F)$ is much larger than the 
kinetic energy term, hence the inclusion of the kinetic energy in Eq.
(\ref{f}) changes the probability amplitude $\psi_F$ only marginally. 
However, inclusion of the kinetic energy in Eq.
(\ref{f})  has the advantage of  leading to a probability amplitude 
$\psi_F$ analytic in space variable $\rho$, whereas the TF approximation 
is not analytic in $\rho$ \cite{fbs2}.

As previously discussed, the Lagrangian (\ref{yy}) with 
(\ref{lagrange-bose}), (\ref{lagrange-fermi}) and 
(\ref{lagrange-bose-fermi}) describes a DBFM under 
axial harmonic confinement of frequency $\omega_z$ and any kind 
of the external potentials $V_B(\rho)$ and $V_F(\rho)$ 
in the cylindric radial directions. 
For our investigation of bosonic vortices in presence of fermions 
we take the following radial traps 
\beq \label{poten}
V_B(\rho )= V_F( \rho) 
= \frac{1}{2} m_B \omega_{\bot}^2 \, \rho^2  \; ,  
\eeq 
as in the study by Modugno {\it et al.} \cite{zzz} and 
Jezek {\it et al.} \cite{jz}, where $\omega_\perp$ refer to the trap 
frequency for bosons. In this 
way the quasi-2D 
mixture is achieved from the so-called 
disk-shaped configuration: the DBFM is confined by an anisotropic 3D 
harmonic potential, $V({\bf r}) =\frac{1}{2}m_B\omega_\bot^2
(\rho^2+\lambda^2 z^2)$, 
where $\lambda =\omega_z/ \omega_{\bot}$ is the 
trap anisotropy. The quasi-2D 
configuration is appropriate for studying vortices in the 
disk-shaped geometry with anisotropy parameter $\lambda \gg 1$. 

%For calculational purpose it is convenient to reduce Eqs. (\ref{e}) and
%(\ref{f}) to dimensionless form by introducing convenient dimensionless
%variables.  In the two experimental situations of Refs. 
%\cite{exp4,exp5}
%$m_B \approx 3m_F$. To keep the algebra simple and without losing
%generality we shall take in the rest of this paper $m_B=3 m_F=
%m({\mbox{Rb}})$, whence $m_R=3m_F/4$, as well as
%$a_{zB}=a_{zF}=a_{zBF}\equiv a_z$. 
 
\section{Numerical result}

The main numerical advantage of working 
with Eqs. (\ref{e}) and (\ref{f}) is that the calculations 
are much faster. In fact, one has to solve the two coupled differential
equations with only one space variable, $\rho$. 
The full 3D problem will require an enormous computational effort. 
In our numerical simulation we consider the $^{40}$K-$^{87}$Rb mixture 
and  take 
$\lambda = 10$,  
$\omega_z = 2\pi \times 
100$ Hz.   We take  $m_B$ as the mass of $^{87}$Rb and $m_F$ as the mass 
of 
$^{40}$K. 
%In the actual calculation  we use dimensionless variables: 
%length in units of $a_{zB}$ and time units of $2/\omega_z$. 
%The unit of length $a_{zB}\approx 1$ $\mu$m  
%and unit of time $2/\omega_z \approx 3$ ms. 
%We also take 
%$a_{zB}=a_{zF}=a_{zBF}$. 

We solve numerically the coupled quantum-hydrodynamic equations 
(\ref{e}) and (\ref{f}) for vortex quantum numbers $l=0$ and $l=1$ 
by using a imaginary-time propagation 
method based on the finite-difference Crank-Nicholson 
discretization scheme elaborated in Ref. \cite{sk1}. 
In this way we obtain the ground-state of the DBFM at a 
fixed value of the vortex quantum number $l$. 
We discretize the quantum-hydrodynamics equations (\ref{e}) 
and (\ref{f}) using time step $0.0003$ ms and space step 
$0.02$ $\mu$m. 

The scattering length $a_{BF}$ is varied from positive 
(repulsive) to negative (attractive) values through zero 
(non-interacting). Note that in the experiments 
the scattering length $a_{BF}$ can be manipulated in 
$^6$Li-$^{23}$Na and $^{40}$K-$^{87}$Rb mixtures near 
recently discovered Feshbach resonances \cite{fesh} 
by varying a background magnetic field. 

\begin{figure}%[!ht]
 
\begin{center}
\includegraphics[width=.8\linewidth]{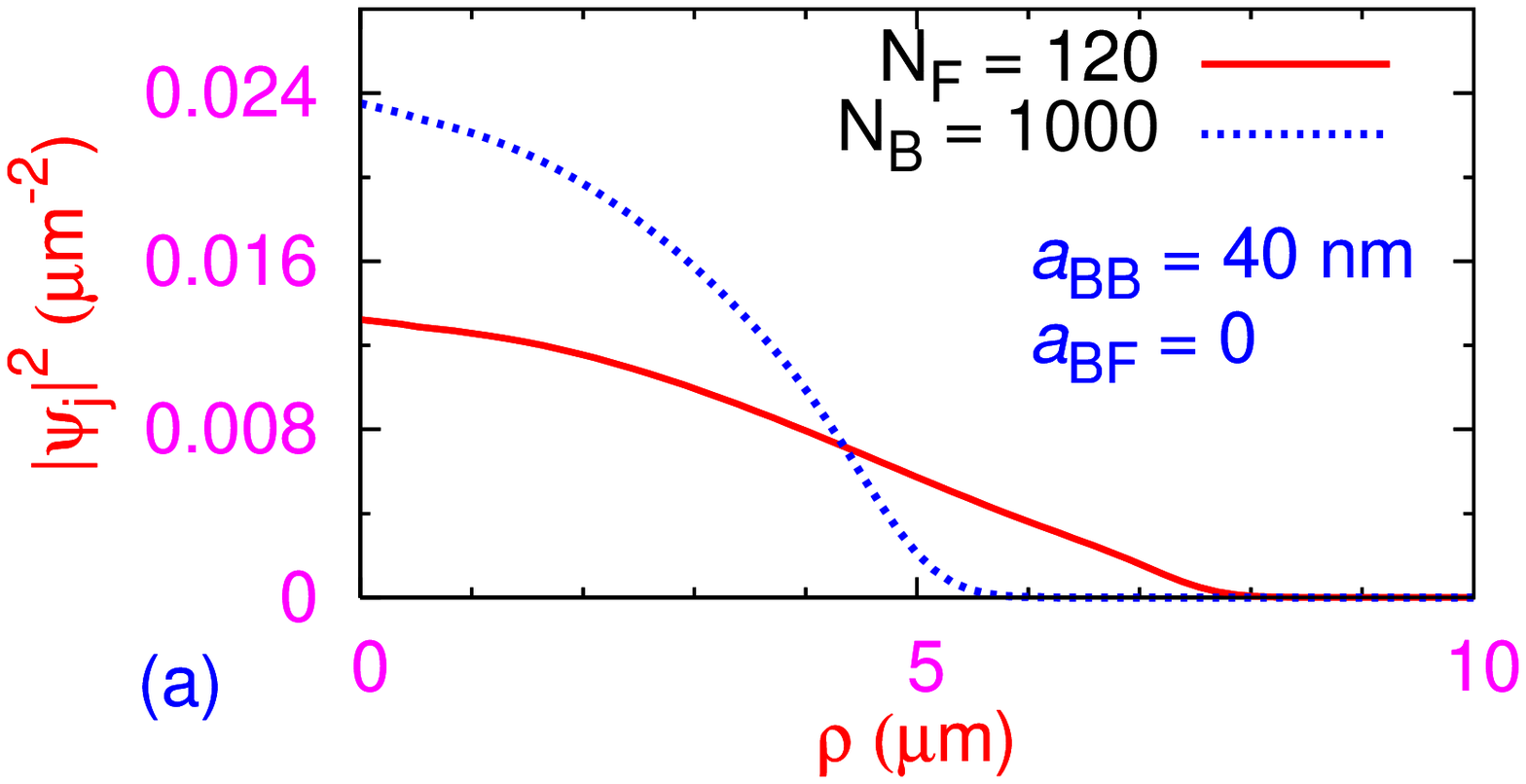}
\includegraphics[width=.8\linewidth]{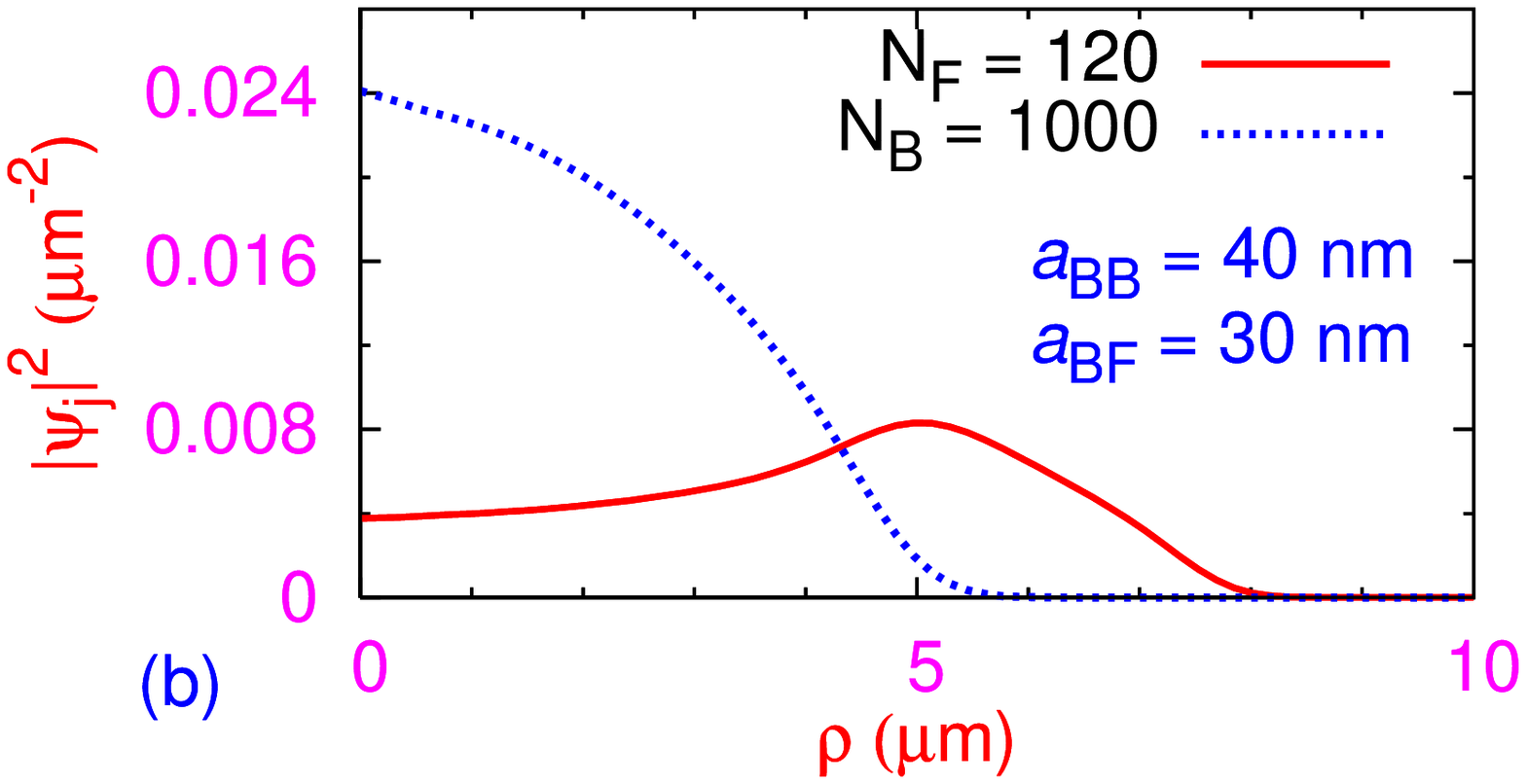}
\includegraphics[width=.8\linewidth]{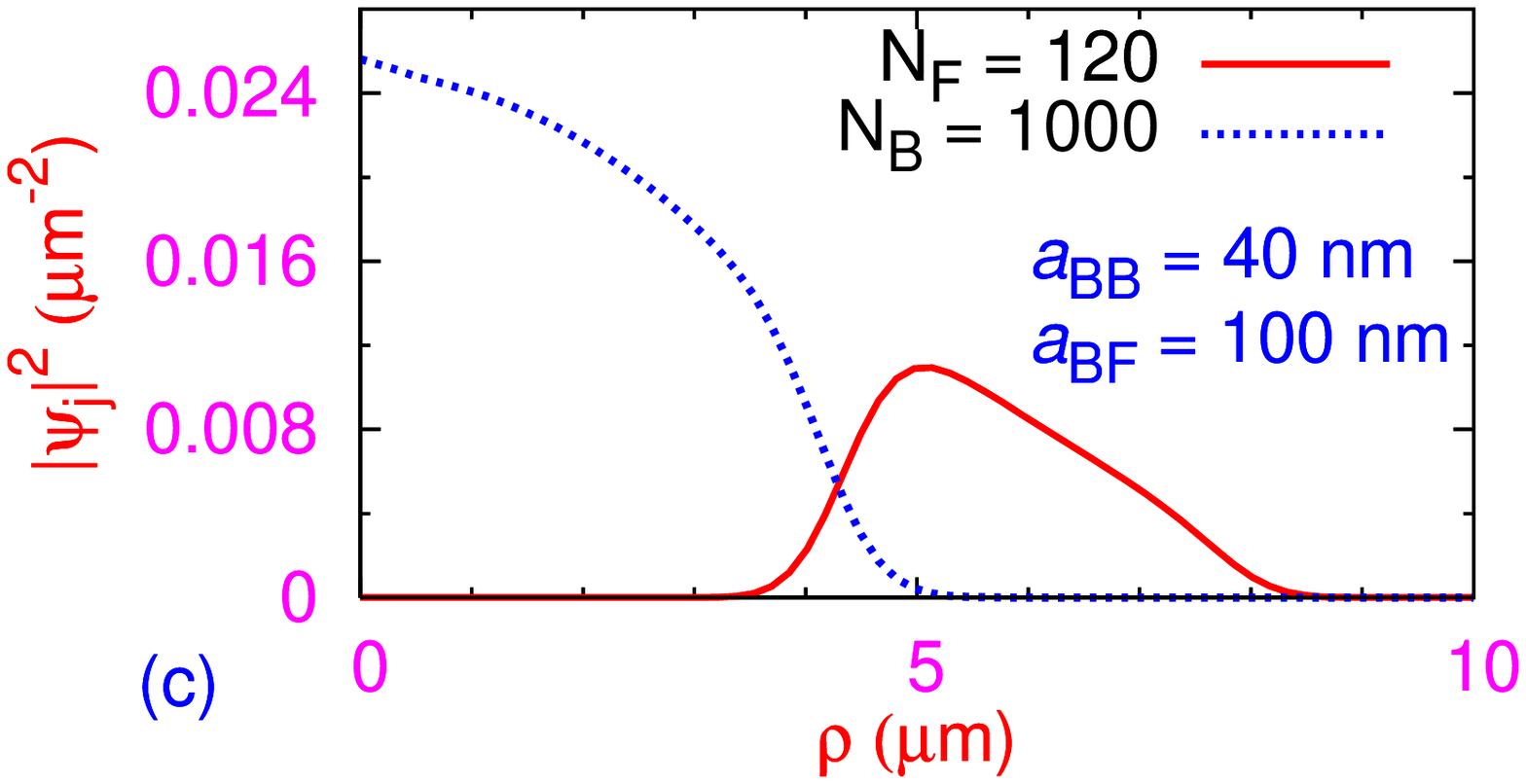}
\includegraphics[width=.8\linewidth]{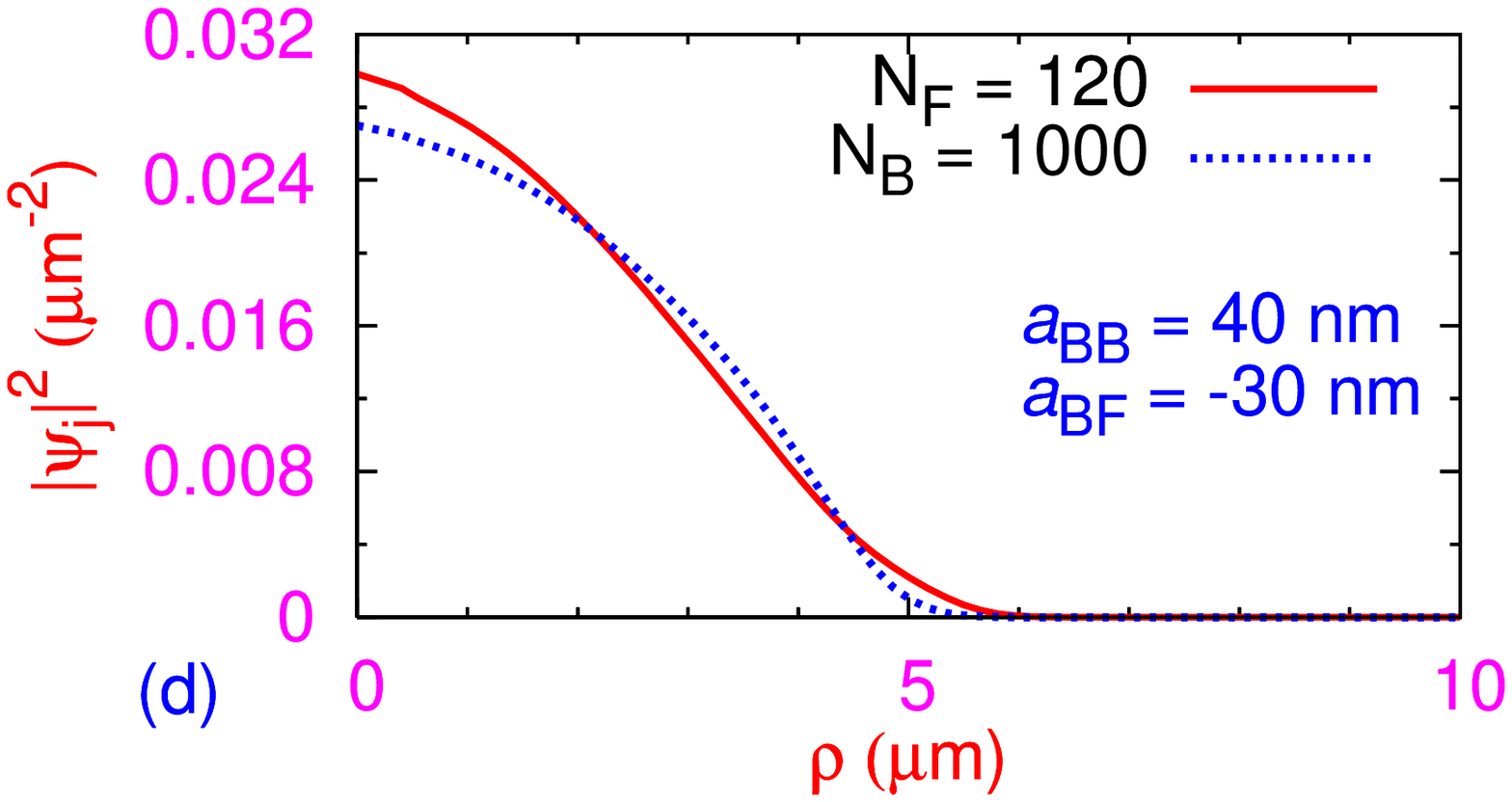}
\end{center}

\caption{(Color online). DBFM with BEC having vortex 
quantum number $l=0$. Probability densities $|\psi_j|^2$ 
of bosons and fermions as a function of the cylindric radial 
coordinate $\rho$.  
$N_B (=1000)$ and $N_F (=120)$ are the numbers of bosons and fermions, 
respectively.  
The trap anisotropy is $\lambda=10$ 
The four panels correspond to different values of the 
Bose-Fermi scattering length $a_{BF} (=0,30$ nm, 100 nm, $-30$ nm) and 
fixed Bose-Bose 
scattering length $a_{BB} (=40$ nm). 
Note that in panel (d) $a_{BF}$ is negative.}
\end{figure} 

\subsection{Mixing-demixing transition}

In the first part of our numerical investigation we consider 
the mixing-demixing transition in the quasi-2D DBFM  
with vortex quantum number $l=0,1$. 
For a sufficiently large repulsive $a_{BF}$ there is 
demixing and for a large attractive $a_{BF}$ there is mixing. If the 
attractive $a_{BF}$ is further increased 
there could be collapse in the DBFM, which we study in detail in the 
next subsection. The mixing-demixing phenomenon is quite similar 
for various boson and fermion numbers and we illustrate it 
choosing $N_F=120$, $N_B=1000$, and $a_{BB}=40$ nm.

\begin{figure}%[!ht] 
\begin{center}
\includegraphics[width=.8\linewidth]{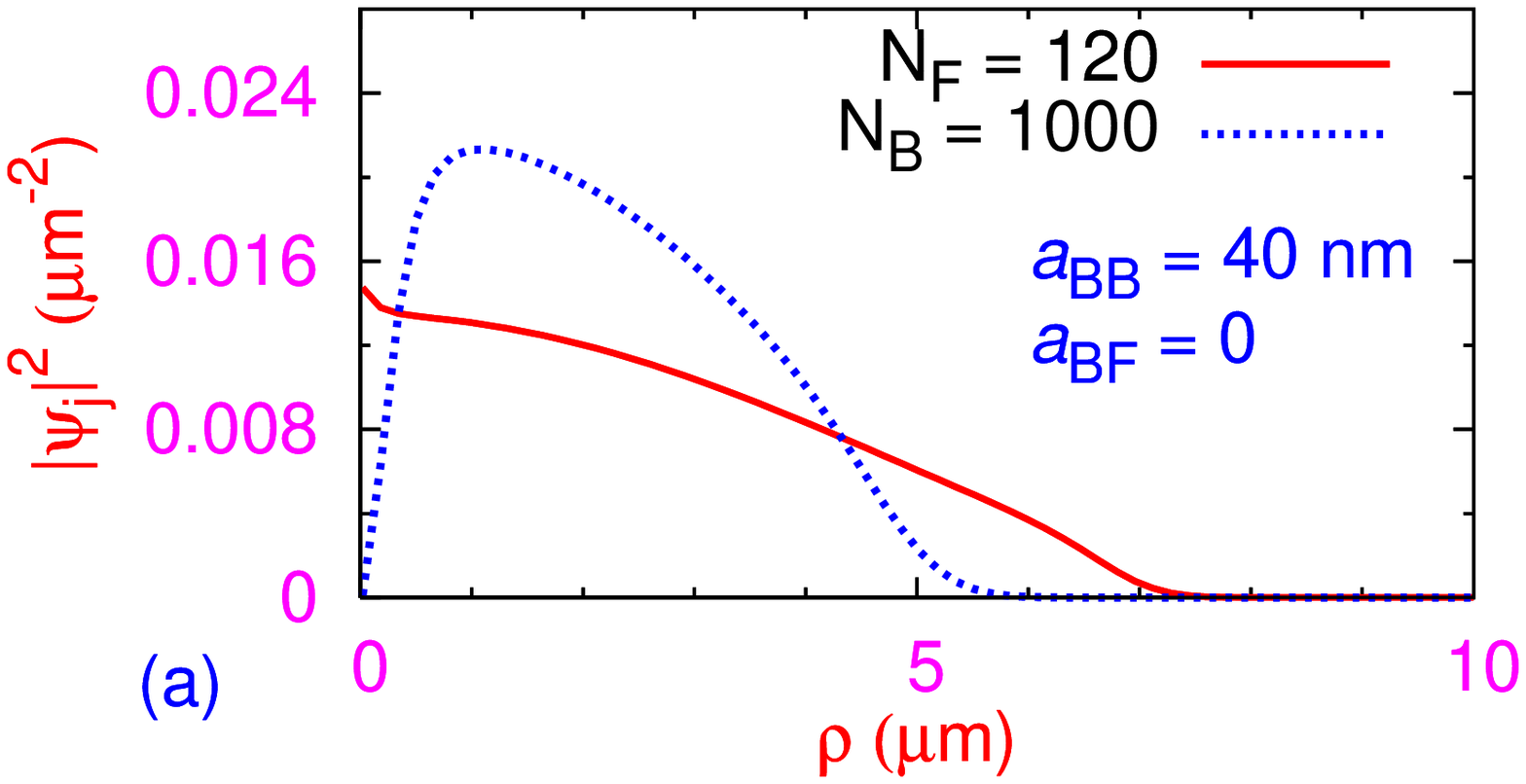}
\includegraphics[width=.8\linewidth]{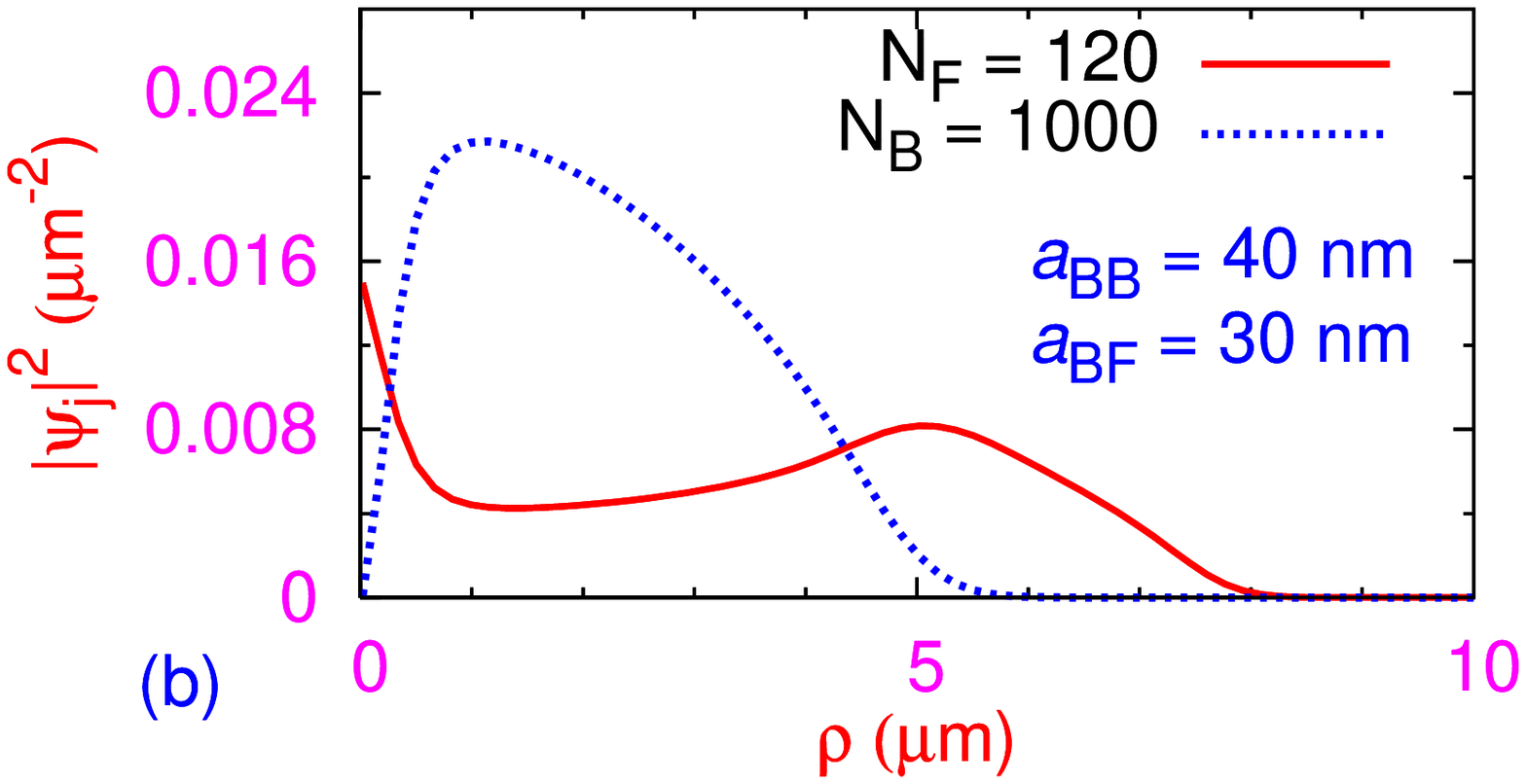}
\includegraphics[width=.8\linewidth]{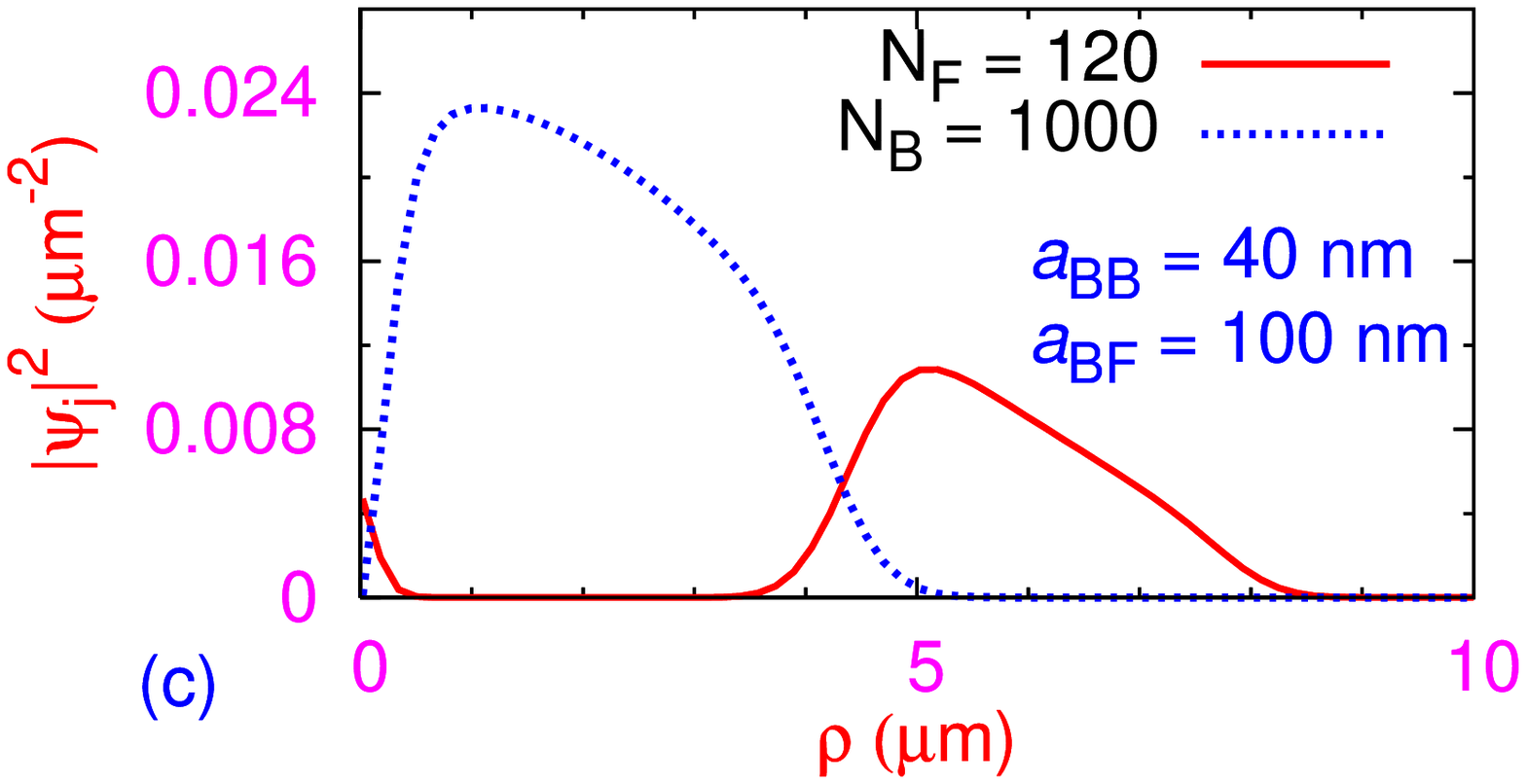}
\includegraphics[width=.8\linewidth]{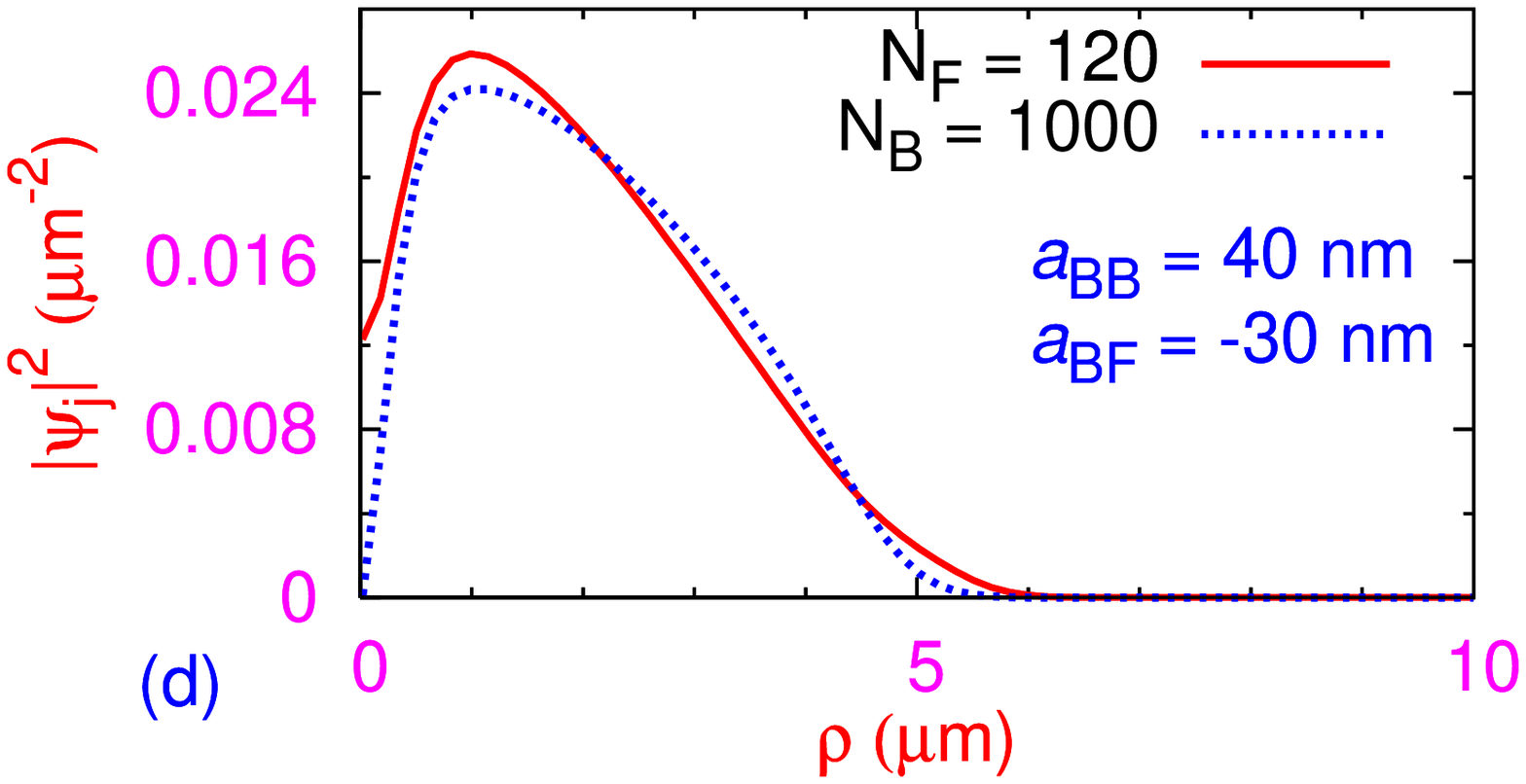}
\end{center} 
\caption{(Color online). 
DBFM with BEC having vortex quantum number $l=1$. 
Probability densities $|\psi_j|^2$
of bosons and fermions as a function of the cylindric radial
coordinate $\rho$. Parameters as in Fig. 1.}
\end{figure}

The results of our imaginary-time calculation with vortex quantum 
number $l=0$ are shown in Fig. 1, where we plot the probability density 
$|\psi_j|^2$ {\it vs.} cylindric radii $\rho$ of the stationary 
boson-fermion mixture in a quasi-2D 
configuration for noninteracting, repulsive and attractive interspecies
interaction. The probability density in Figs. 1 and 2 is 
normalized to unity:
$2\pi \int_0^\infty |\psi_j|^2 \rho d \rho =1$. 
In all cases, with $a_{BF}>0$, because of the large nonlinear 
Pauli-blocking fermionic repulsion, the fermionic profile 
extends over a larger region of space than the bosonic one. 
As shown in Fig. 1, in agreement with previous studies in the $l=0$ state,  
a complete mixing-demixing transition is found 
by increasing $a_{BF}$ from $a_{BF} = 0$ to $a_{BF} = 100$ nm. 
Instead, in the case of attractive boson-fermion interaction ($a_{BF}<0$)  
we find that the fermionic cloud is pulled inside 
the bosonic one and a complete overlap between 
the two clouds is then achieved. 
With further increase in boson-fermion interaction 
the system collapses. In Fig. 1(d), where $a_{BF} = -30$ nm, 
just below the critical value for collapse, we find 
an almost complete mixing between the bosonic and fermionic components. 

\begin{figure}%[!ht]
 
\begin{center}
\includegraphics[width=1.\linewidth]{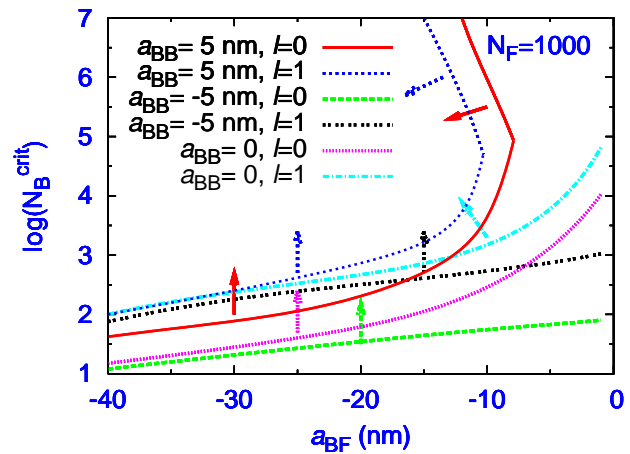}
\end{center}

\caption{(Color online). Critical number $N_B^{\mbox{crit}}$ 
of bosons {\it vs.} Bose-Fermi scattering length $a_{BF}$ 
in the DBFM. The number $N_F$ of fermions is fixed at 1000. 
Vortex quantum number $l$ and Bose-Bose scattering length 
$a_{BB}$ are instead varied. 
The region of collapse is indicated by arrows.}
\end{figure}

In the second part of the investigation 
we consider a BEC with vortex quantum number $l=1$ 
in a DBFM with the same parameters of Fig. 1. 
The results 
are displayed in Fig. 2. The non-interacting case ($a_{BF}=0$)
is exhibited in Fig. 2(a) with $N_F=120$, 
$N_B=1000$, $a_{BB}=40$ nm. The fermionic profile in this case 
is quite similar to that in the $l=0$ state 
exhibited in Fig. 1(a). However, the bosonic profile has developed
a dip near origin due to the $l=0$ vortex state. In Fig. 2(b) upon 
introducing a interspecies repulsion between bosons and fermions a 
demixing has started and the fermionic wave function is partially pushed
out from the central region for $a_{BF}=30$ nm. 
This demixing has increased in Fig. 2(c) for $a_{BF}=100$ nm. 
The fermionic profile in Fig. 2(c) for the vortex state with $l=1$ is 
quite similar to the corresponding state in Fig. 1(c) for $l=0$. 
Finally, we find that an attractive boson-fermion interaction increases
the mixing of boson and fermion components and the mixing is maximum for a
critical value of $a_{BF}$ before the occurrence of collapse in the
DBFM. The boson and fermion profiles for $a_{BF} = -30$ 
nm just below the threshold for collapse is shown in Fig. 2(d). In this
case the mixing is so perfect that the fermionic profile has developed a
central dip near $\rho =0$ reminiscent of a vortex state as in the bosonic
component. However, near $\rho= 0$ the fermionic wave function 
$\psi_F(\rho)$ tends to a constant value and does not 
have the vortex state behavior $\psi_B(\rho) \sim \rho^l$. 
This shows that the fermionic state is not really a vortex 
but due to mixing it tends to simulate one. 

\subsection{Stability and Collapse}

For given values of $N_B$, $N_F$, $a_{BB}$ and angular momentum 
$l$, always a stable configuration is achieved for a repulsive 
(positive) $a_{BF}$. However, for a sufficiently large attractive (negative) 
$a_{BF}$, the system collapses as the overall attractive interaction 
between bosons and fermions supersedes  the overall    
stabilizing repulsion of the system thus leading to instability. 
Also, alternatively, for a fixed $a_{BF}$, $N_F$, $a_{BB}$ and 
angular momentum $l$, the system may collapse for $N_B$ greater 
than a critical value $N_B^{\mbox{crit}}$.  
This may happen for all values of the parameters and 
a typical situation is illustrated in Fig. 3, where 
we plot $N_B^{\mbox{crit}}$ {\it vs.} $a_{BF}$ in different 
cases for $N_F=1000$. One can have collapse in a single or both 
components. After collapse the radius of the system reduces to a very 
small value.

For $a_{BB} \le 0$ the system is stable for 
$N< N_B^{\mbox{crit}}$ for both $l=0$ and 1 as one can see 
from Fig. 3, where we plot results for $a_{BB}=0$ and $-5$ nm. 
The region of instability and collapse is indicated by 
arrows for $N> N_B^{\mbox{crit}}$.
In Fig. 3 we find that the region of stability has increased after the 
inclusion of the angular momentum term. 
Due to the stabilizing repulsive centrifugal 
term $\hbar^2 l^2 /(2 m_B \rho^2)$ the rotating ($l=1$) DBFM is more  
stable than the nonrotating ($l=0$) one \cite{sk3}. 

For $a_{BB}>0$, the bosons have a  net repulsive energy and the 
system does not collapse unless the strength 
of boson-fermion attraction $|a_{BF}|$ is increased 
beyond a critical value. In this situation an 
interesting scenario appears at fixed $a_{BF}$, $a_{BB}$, $N_F$ and 
$l$, as $N_B$ is increased. Increasing $N_B$ from a small value
past the critical value $N_B^{\mbox{crit}}$, the system collapses 
because the attractive interaction in the Lagrangian density 
(\ref{lagrange-bose-fermi}) becomes large 
enough to overcome the stabilizing repulsions in boson-boson and 
fermion-fermion subsystems. However, when $N_B$ becomes very large 
($N_B\gg N_F$) past a second critical value, the
attractive interaction in the Lagrangian density 
(\ref{lagrange-bose-fermi}) will  become 
small compared to the overall repulsion of the system and a stable 
configuration can again be obtained. This is clearly illustrated 
in Fig. 5. Note that, in Eq. (\ref{lagrange-bose-fermi}),
$|\psi_B|^2\propto N_B$ and $|\psi_F|^2\propto N_F$ 
and for a fixed total number of atoms $(N_B+N_F)$
${\cal L}_{BF}$ becomes  large for $N_B \approx N_F$ and small for 
$N_B \gg N_F$ or $N_B \ll N_F$.
Hence, a small number $N_F$ of fermions should not destabilize the 
stable configuration of a large number $N_B$ of repulsive bosons. 
Similarly, a small number $N_B$ of bosons should not destabilize the
stable configuration of a large number $N_F$ 
of repulsive fermions. This feature, that is explicitly 
shown in Fig. 3, should be quite general independent of 
dimensionality of space. It is not clear why this feature was not found 
in the 3D theoretical study of Ref. \cite{yyy}. 

\begin{figure}%[!ht]
 
\begin{center}
\includegraphics[width=1.\linewidth]{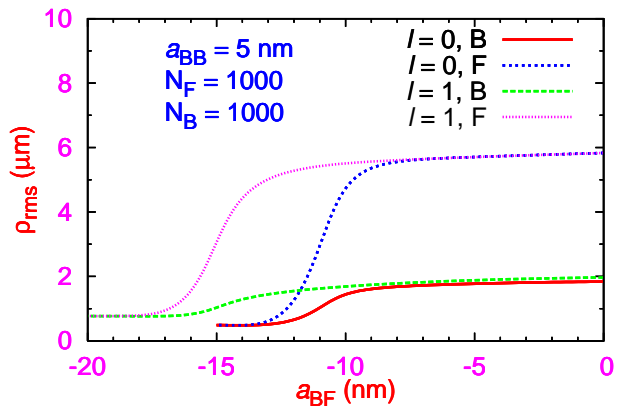}
\end{center}

\caption{(Color online). 
Root mean square radius $\rho_{\mbox{rms}}$ of the 
two clouds in the DBFM as a function of the Bose-Fermi scattering 
length $a_{BF}$. Here the number $N_B$ 
of bosons (labeled B) is equal to the number 
$N_F$ of fermions (labeled F) fixed at 1000. The results are shown for 
two values 
of the BEC vortex quantum number $l=0,1$.}

\end{figure}

Next we analyze how the system moves towards collapse as the parameters of 
the model are changed. First we consider the passage to collapse as the 
attractive strength of boson-fermion interaction is increased.  To see 
this we plot the root-mean-square (rms) radii $\rho_{\mbox{rms}}$  of 
bosons and fermions vs. 
$a_{BF}$ in several cases in Fig. 4. The rms radii remain fairly 
constant away from the region of collapse. However, as $a_{BF}$
approaches the value for collapse the rms radii decrease rapidly to a 
small value signalling the collapse. In this case both bosons and 
fermions experience collapse simultaneously. 

We also studied the collapse 
for a fixed attractive $a_{BF}$, $a_{BB}$ and $N_F$, 
while $N_B$ is varied,  
to demonstrate that  stable configuration can be attained 
simultaneously for 
bosons and fermions for small and large $N_B$, e.g. for $N_B\ll N_F$ or 
$N_B\gg N_F$. For intermediate $N_B$ 
there is collapse in 
either  bosons or fermions or both. This is illustrated in Fig. 5 where 
we plot $\rho_{\mbox{rms}}$ vs. $\log(N_B)$ for $N_F=1000$ and  
$a_{BB}=5$ nm for both bosons and fermions for $l=0$ and 1. In both 
cases ($l=0,1$), as $N_B$ is increased from a small value,  the collapse 
is initiated near $N_B= N_B^{\mbox{crit}}\approx 1000$ when 
the radii of both bosons and fermions suddenly drop  to a small value 
signalling a collapse in both subsystems. With further  
increase in $N_B$ near $N_B\approx 8000$
the bosons pass to a stable state from a 
collapsed state and the corresponding rms radii increase with $N_B$. 
The fermions continue in a collapsed state  with a small 
radii. However, near $N_B\approx  5\times 10^6$ the fermions 
also come out of the collapsed state and 
with the increase of $N_B$ the fermion radius starts 
to increase. For $N_B > 10^{7}$, a stable configuration of the DBFM is 
obtained as the boson-boson and fermion-fermion repulsion compensates 
for the boson-fermion attraction.   

\begin{figure}%[!ht]
 
\begin{center}
\includegraphics[width=1.\linewidth]{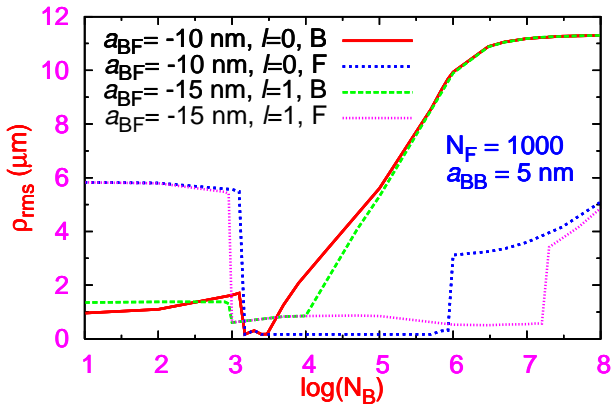}
\end{center}

\caption{(Color online). Root mean square radius $\rho_{\mbox{rms}}$ of 
the
two clouds in the DBFM as a function of $\log(N_B)$, 
that is the logarithm of the number $N_B$ of bosons (labeled B). 
Here the number $N_F$ of fermions (labeled F) is fixed at 1000. 
The results are show for two values of the BEC vortex 
quantum number $l=0,1$.}
\end{figure}

\section{Conclusion} 
 
We have used a coupled set of quantum-hydrodynamic 
equations to study the mixing-demixing transition 
as well as stability and collapse 
of a trapped DBFM. 
The model equations are solved by 
imaginary time propagation of the finite-difference 
Crank-Nicholson algorithm. In our analysis the 
Bose-Einstein condensate is strictly 2D while 
the Fermi gas is not: for this reason we have introduced 
a new fermionic density functional which accurately takes into 
account the dimensional crossover of fermions from 2D to 3D.  
In the study of the mixing-demixing transition 
we have taken the boson-boson interaction to be 
repulsive and the boson-fermion interaction
to be both attractive and repulsive. 
By considering a bosonic vortex with quantum number $l$,  
we have found that in both $l=0$ and $l=1$ cases 
the mixing could be almost complete up to the critical value 
of boson-fermion attraction beyond which the 
system collapses. When the boson-fermion interaction 
is turned repulsive, there is the mixing-demixing transition 
which is regulated by the boson-fermion repulsion. 

We have also studied the collapse 
in $l=0$ and $l=1$ cases. The $l=1$ system is found to be more 
stable due to the centrifugal kinetic term. 
We have investigated in detail how stability is affected 
when the boson-boson and boson-fermion interaction 
as well as boson and fermion numbers are varied. 

The present analysis is based on mean-field Eqs. (\ref{e}) and (\ref{f}) 
for the Bose-Fermi mixture, 
which are very similar in structure to those satisfied by a Bose-Bose 
mixture \cite{bbcol}. In the mean-field equations for a Bose-Bose 
mixture 
all nonlinearities are 
cubic in nature. In the present mean-field equations for a  Bose-Fermi 
mixture apart from the intraspecies Fermi (diagonal) nonlinearity 
arising from the 
Pauli principle in Eq.  (\ref{f}), given by Eq. (\ref{chem-fermi}), all 
other 
nonlinearities are also cubic in nature. In Eq. (\ref{chem-fermi}) the 
nonlinearity is partly cubic and partly has a different form; whereas in 
the strict 2D limit this nonlinearity is entirely cubic in nature.  
Bearing such a similarity with the mean-field equations of the Bose-Bose 
mixture, the $l=0$ results for mixing-demixing and collapse in 
 Bose-Fermi mixture presented here are   
expected to be similar to those of a Bose-Bose mixture provided that the 
scattering lengths and trap parameters are adjusted in two cases to lead 
to similar strengths of the nonlinearities. [It has been 
demonstrated, similar to the present Bose-Fermi mixture,  that a 
Bose-Bose mixture with intraspecies repulsion and interspecies 
attraction can experience collapse \cite{bbcol}.] But the present $l=1$ 
results with a bosonic vortex in a Bose-Fermi mixture have no analogy 
with in the Bose-Bose case. A slowly rotating Bose-Fermi mixture can 
have a 
quantized bosonic vortex with $l=1$ with no vortex in the fermions; 
whereas a 
similar Bose-Bose mixture  should have a $l=1$ vortex in both the 
bosonic
components in a stable stationary configuration. Consequently, the 
mean-field equations satisfied by the 
slowly rotating Bose-Fermi mixture will be distinct from those satisfied 
by a slowly rotating Bose-Bose mixture and the present results for $l=1$ 
should be distinct from those for a Bose-Bose mixture.

The present findings can be verified 
in experiments on DBFMs, 
specially for the vortex state, thus presenting 
yet another critical test of our quantum-hydrodynamic model.
  
\acknowledgments
L.S. thanks Flavio Toigo for useful discussions. 
The work of S.K.A. is supported in part by the CNPq and FAPESP of 
Brazil. 

\section*{Appendix} 

In this appendix we derive the zero-temperature equations of state 
for an ideal Fermi gas that is uniform in the cylindric 
radial direction but under harmonic confinement 
in the cylindric axial direction. In particular we obtain 
the chemical potential $\mu_F$ and the energy density ${\cal E}_F$ 
as a function of the 2D uniform radial density $n_F$ of the Fermi gas.  
\par 
Let us consider an ideal Fermi gas in a box of length $L$ 
along $x$ and $y$ axis and harmonic potential of frequency $\omega_z$ 
along the $z$ axis. The total number of particles is 
\beq 
N_F = \sum_{i_x i_y i_z} 
\theta( \mu - \epsilon_{i_x i_y i_z} ) \; , 
\eeq
where the single particle energy reads 
\beq
\epsilon_{i_x i_y i_z} = {\hbar^2 \over 2m_F} 
{(2\pi)^2 \over L^2} (i_x^2 + i_y^2) 
+ \hbar \omega_z (i_z + {1\over 2}) \; .   
\eeq
Here $i_x$, $i_y$ are integer quantum numbers 
and $i_z$ is a natural quantum number. 
Let us approximate $i_x$ and $i_y$ by real numbers (see also 
\cite{sala-ideal}. Then 
\beq
N_F = \sum_{i_z=0}^{\infty} \int di_x di_y \, 
\theta( \mu - \epsilon_{i_x i_y i_z} ) \; . 
\eeq
Setting $k_x = {2\pi\over L} i_x$ and 
$k_y = {2\pi\over L} i_y$, the number of particles can be rewritten as 
\beq
N_F = \sum_{i_z=0}^{\infty} {L^2\over (2\pi)^2} 
\int dk_x dk_y \, \theta( \mu - 
\epsilon_{k_x k_y i_z} ) \; .
\eeq
The 2D density is then 
\beq
n_F = {N_F\over L^2} = {1\over 4\pi^2} 
\sum_{i_z=0}^{\infty} \int dk_x dk_y \, 
\theta( \mu - \epsilon_{k_x k_y i_z} ) \; . 
\eeq
Now we re-write the density $n_F$ in the following way 
\begin{eqnarray} \label{xy}
n_F &=& {1\over 4\pi^2} 
\int dk_x dk_y\, \theta( \mu - \epsilon_{k_x k_y 0} ) 
\nonumber \\
&+& {1\over 4\pi^2} \sum_{i_z=1}^{\infty} \int dk_x dk_y\, 
\theta( \mu - \epsilon_{k_x k_y i_z} ) \; . 
\end{eqnarray}
The first term is the density of a strictly 
2D Fermi gas (only the lowest single-particle mode 
along the $z$ axis is occupied) and the second term 
is the density which takes into account of all single-particle 
modes along the $z$ axis, apart the lowest one. 
Setting $k^2 = k_x^2 + k_y^2$ the first term 
of the density $n_F$ can be written as 
\begin{eqnarray} 
{1\over 4\pi^2} \int 2\pi k dk\, \theta( \mu - {\hbar^2 k^2 \over 2m} - 
{1\over 2}\hbar \omega_z) 
= {1\over 2\pi a_{zF}^2} 
\left( {\mu\over \hbar \omega_z} - 
{1\over 2} \right) , \nonumber  
\end{eqnarray}
where $a_{zF}=\sqrt{ \hbar /(m_F\omega_z)}$. 
The second term of the 2D density $n_F$ is evaluated by transforming it 
to an integral and is
given by 
\begin{eqnarray} 
\sum_{i_z=1}^{{\mu\over \hbar\omega_z} - {1\over 2}} 
{1\over 2 \pi a_{zF}^2} \, 
\left({\mu\over \hbar\omega_z} - (i_z+{1\over 2}) \right) 
= {1\over 4 \pi a_{zF}^2} 
\left({\mu\over \hbar \omega_z} - {3\over 2}\right)^2  
\; . \nonumber
\end{eqnarray}

In conclusion the 2D density $n_F$ is given by 
\begin{eqnarray}
n_F = {1\over 2 \pi a_{zF}^2 } \left[ 
\left({\mu_F\over \hbar \omega_z} \right) + 
{1\over 2} \left({\mu_F\over \hbar \omega_z} - 1\right)^2 
\theta({\mu_F\over \hbar \omega_z}-1) \right] \; ,  
\label{n2d-mu}
\end{eqnarray}
% \bot changed to z
where $\mu_F=\mu - \hbar \omega_{z}/2$ is the chemical potential
minus the ground-state harmonic energy along the $z$ axis. 

The Eq. (\ref{n2d-mu}) can be easily inverted and we find 
\beq 
\mu_F = \hbar \omega_z 
\left\{ \ba{ll}
2\pi n_F a_{zF}^2 & \mbox{ for  $0\le n_F a_{zF}^2 < {1\over 2\pi}$} \\
\sqrt{4\pi n_F a_{zF}^2 - 1} 
& \mbox{ for  $n_F a_{zF}^2 \ge {1\over 2\pi}$} \, .
\ea \right. 
\label{chem-fermi} 
\eeq
Equation (\ref{chem-fermi}) carries the fermionic nonpolynomial 
nonlinearity in quasi-2D formulation to be used in Eqs. (\ref{e}) and 
(\ref{f}). In the strict 
2D limit, the second term in Eq. (\ref{xy}) is absent and $\mu_F=2\pi 
\hbar\omega_z n_F a_{zF}^2$ corresponding to a cubic  nonlinearity.

We can also derive the energy 
density ${\cal E}_F$ of the Fermi gas from the following 
formula of zero-temperature thermodynamics 
\beq 
{\cal E}_F = \int \, dn_F \, \mu_F(n_F) \, .  
\eeq
In this way we get 
\beq
{\cal E}_F  = {\hbar \omega_z\over a_{zF}^2} 
\left\{ \ba{ll} 
\pi (n_F a_{zF}^2)^2 \quad  
\mbox{for} \quad 0\le n_F a_{zF}^2 < {1\over 2\pi} \\
{1\over 6\pi}(4\pi n_F a_{zF}^2 - 1)^{3/2} + {1\over 12 \pi}
\quad \mbox{for} \quad n_F a_{zF}^2 \ge {1\over 2\pi} \, . 
\ea \right.
\label{energy-fermi} 
\eeq
The Fermi gas is strictly 2D only for  
$0\le n_F <{1/(2\pi a_{zF}^2)}$, i.e. for 
$0\le \mu_F < \hbar \omega_z$. 
For $n_F >{1/(2\pi a_{zF}^2)}$, i.e. for $\mu_F> \hbar \omega_z$,  
several single-particle states of the harmonic oscillator
along the $z$ axis are occupied and the gas has the 2D-3D crossover. 
Finally, for $n_F \gg {1/(2\pi a_{zF}^2)}$, i.e. for $\mu_F 
\gg \hbar \omega_z$, the Fermi gas becomes 3D \cite{sala-ideal}. 
\par 
The equations of state (\ref{chem-fermi}) and (\ref{energy-fermi}) 
can be used to write down, in the local density approximation, 
the density functionals of the quasi-2D Fermi gas in presence 
of an additional external potential $V(\rho)$ in the cylindric radial 
direction $\rho$. In this case the 2D fermionic density $n_F$ becomes 
a function of the radial coordinate: $n_F=n_F(\rho)$.


\begin{thebibliography}{99}

\bibitem{exp1} B. DeMarco and D. S. Jin, Science {\bf 285}, 1703
(1999).

\bibitem{exp2} K. M. O'Hara, S. L. Hemmer, M. E. Gehm,  S. R. Granade, and
J. E. Thomas, Science {\bf 298}, 2179 (2002).

\bibitem{exp3} F. Schreck, L. Khaykovich, K. L. Corwin, G. Ferrari,
T. Bourdel, J. Cubizolles,  and C. Salomon,
Phys. Rev. Lett. {\bf 87}, 080403 (2001);
A. G. Truscott, K. E. Strecker, W. I. McAlexander,
G. B. Partridge, and R. G. Hulet, Science {\bf 291}, 2570 (2001). 

\bibitem{exp4} Z. Hadzibabic, C. A. Stan,
K. Dieckmann, S. Gupta, M. W. Zwierlein, A. Gorlitz, and W. Ketterle,
Phys. Rev. Lett. {\bf 88}, 160401 (2002).

\bibitem{exp5} G. Modugno, G. Roati, F. Riboli, F. Ferlaino, R. J. Brecha,
and M. Inguscio, Science {\bf 297}, 2240 (2002). 

\bibitem{exp5x} G. Roati,  F. Riboli, G. Modugno, and M. Inguscio,
Phys. Rev. Lett. {\bf 89}, 150403 (2002).

\bibitem{exp6} K. E. Strecker, G. B. Partridge, and R. G. Hulet,
Phys. Rev. Lett. {\bf 91}, 080406 (2003); Z. Hadzibabic, 
S. Gupta, C. A. Stan, C. H. Schunck,  M. W. Zwierlein,
K. Dieckmann, and W. Ketterle, {\it ibid.} {\bf 91}, 160401 (2003).

\bibitem{yyy} R. Roth, Phys. Rev. A {\bf 66}, 013614 (2002).

\bibitem{yyy1} K. Molmer, Phys. Rev. Lett.
{\bf 80}, 1804 (1998).

\bibitem{yyy2} R. Roth and 
H. Feldmeier,  Phys. Rev. A  {\bf 65}, 021603(R) (2002); T. Miyakawa,
T. Suzuki, and H. Yabu,  {\it ibid.}  {\bf 64}, 033611 (2001).

\bibitem{md1} Y. Takeuchi and H. Mori,  Phys. Rev. A {\bf 72},
063617 (2005).

\bibitem{md2} Z. Akdeniz, A. Minguzzi, P. Vignolo, and 
M. P. Tosi, Phys. Lett. A {\bf 331}, 258 (2004); 
P. Capuzzi, A. Minguzzi, and M. P. Tosi, Phys. Rev. A 
{\bf 68}, 033605 (2003).

\bibitem{zzz} M. Modugno, F. Ferlaino, F. Riboli, G. Roati, 
G. Modugno, and M. Inguscio, 
Phys. Rev. A {\bf 68}, 043626 (2003);           
X.-J. Liu, M. Modugno, and H. Hu, {\it ibid.}  {\bf 68}, 053605 (2003).  

\bibitem{capu}
P. Capuzzi, A. Minguzzi, and M. P. Tosi, 
Phys. Rev. A {\bf 69}, 053615 (2004); 
{\bf 67}, 053605 (2003). 

\bibitem{anna-minguzzi} 
A. Minguzzi, P. Vignolo, M. L. Chiofalo, and M. P. Tosi, 
Phys. Rev. A {\bf 64}, 033605 (2001). 

\bibitem{ska} S. K. Adhikari, Phys. Rev. A {\bf 70}, 043617 (2004). 

\bibitem{skac}S. K. Adhikari, New J. Phys. {\bf 8}, 258 (2006). 
 
\bibitem{fesh} C. A. Stan, M. W. Zwierlein, C. H. Schunck,
S. M. F. Raupach, and W. Ketterle, Phys. Rev. Lett. {\bf 93}, 143001
(2004); S. Inouye, J. Goldwin, M. L. Olsen, C. Ticknor, 
J. L. Bohn, and D. S. Jin, {\it ibid.}
{\bf 93}, 183201 (2004).

\bibitem{mdmff}S. K. Adhikari, Phys. Rev. A {\bf 73}, 043619 (2006); S. 
K. Adhikari and B. A. Malomed, {\it ibid.} 
 {\bf 74}, 053620 (2006).


\bibitem{bongs} C. Ospelkaus, S. Ospelkaus, K. Sengstock, and 
K. Bongs, Phys. Rev. Lett. {\bf 96}, 020401 (2006).     

\bibitem{ccol} M. Zaccanti, C. D'Errico,  F. Ferlaino,  G. Roati,
M. Inguscio,  and  G.  Modugno, Phys. Rev. A {\bf 74}, 041605(R) 
(2006); S.  Ospelkaus, C. Ospelkaus, L.
Humbert, K.  Sengstock, and K.  Bongs, Phys. Rev. Lett. {\bf 97},  
120403 (2006).

\bibitem{donn} R. J. Donnelly, {\it Quantized Vortices in Helium II}
(Cambridge University Press, Cambridge, 1991).

\bibitem{skal} S. K. Adhikari, Am. J. Phys. {\bf 54}, 362
(1986).

\bibitem{exptv} K. W. Madison, F. Chevy, W. Wohlleben, and J. Dalibard, 
Phys. Rev. Lett. {\bf 84}, 806 (2000);
M. R. Matthews, B. P. Anderson, P. C. Haljan, D. S. Hall, 
C. E. Wieman, and E. A. Cornell,
{\it ibid.} {\bf 83}, 2498 (1999).

\bibitem{d1} D. L. Feder, C. W. Clark, and B. I. Schneider, 
Phys. Rev. Lett. {\bf 82}, 4956 (1999).

\bibitem{d2} B. Jackson, J. F. McCann, and C. S. Adams, 
Phys. Rev. A {\bf 61}, 013604 (1999).
\bibitem{d3} R. J. Marshall, G. H. C. New, K. Burnett, and S. Choi, 
Phys. Rev. A {\bf 59}, 2085 (1999).
\bibitem{d4}L. Dobrek, M. Gajda, M. Lewenstein, K. Sengstock, G. Birkl,
and W. Ertmer, Phys. Rev. A {\bf 60}, R3381 (1999).    

\bibitem{d5} A. A. Svidzinsky and 
A. L. Fetter, Phys. Rev. A {\bf 62}, 063617 (2000);
A. L. Fetter and A. A. Svidzinsky, J. Phys.: Condens. Matter
{\bf 13}, R135 (2001). 

\bibitem{rokhsar} D.S. Rokhsar, Phys. Rev. Lett. {\bf 79}, 2164 (1997). 

\bibitem{sala-toro}
L. Salasnich, A. Parola and L. Reatto,
Phys. Rev. A {\bf 59}, 2990 (1999);
A. Parola, L. Salasnich, R. Rota, and L. Reatto, {\it ibid} {\bf 72}, 
063612 (2005);
L. Salasnich, A. Parola and L. Reatto, {\it ibid} {\bf 74}, 031603 
(2006). 

\bibitem{dalfovo-toro} M. Modugno, C. Tozzo, F. Dalfovo, 
Phys. Rev. A {\bf 74} 061601(R) (2006); S. Schwartz, M. Cozzini, 
C. Menotti, I. Carusotto, P. Bouyer, S. Stringari, 
New J. Phys. {\bf 8}, 162 (2006). 



\bibitem{karx} T. Karpiuk, M. Brewczyk, and K. Rzazewski,
J. Phys. B {\bf
35}, L315
(2002);  J. Phys.
B {\bf 36}, L69 (2003).



\bibitem{sala-solitonBF} L. Salasnich, S. K. Adhikari, and F. Toigo,  
Phys. Rev. A  {\bf 75}, 023616 (2007). 

\bibitem{fbs2} S. K. Adhikari,  Phys. Rev. A {\bf 72}, 053608 (2005).

\bibitem{fds} S. K. Adhikari,  J. Phys. B  {\bf 38}, 3607 (2005); Laser 
Phys. Lett. {\bf 3}, 605 (2006).

\bibitem{bongs1} T. Karpiuk, K. Brewczyk, S. Ospelkaus-Schwarzer, 
K. Bongs, M. Gajda, and K. Rzazewski, 
Phys. Rev. Lett. {\bf 93}, 100401 (2004). 

\bibitem{stringari-book} L. P. Pitaevskii and S. Stringari, 
{\it Bose-Einstein Condensation} 
(Oxford University Press, Oxford 2003);
F. Dalfovo, S. Giorgini, L. P. Pitaevskii, and S. Stringari,
Rev. Mod. Phys. {\bf 71}, 463 (1999); V. I. Yukalov, Laser Phys. Lett. 
{\bf 1}, 435 (2004);  V. I. Yukalov and M. D. Girardeau, {\it ibid.} 
{\bf 2}, 375 (2005).   

\bibitem{lipparini} E. Lipparini, {\it Modern Many-Particle Physics: 
Atomic Gases, Quantum Dots and Quantum Fluids} 
(World Scientific, Singapore, 2003). 

\bibitem{sala-npse} L. Salasnich, Laser Phys. {\bf 12}, 198 (2002);
L. Salasnich, A. Parola, and L. Reatto,
Phys. Rev. A {\bf 65}, 043614 (2002). 

\bibitem{sala-3DBEC-1DTG} L. Salasnich, A. Parola, L. Reatto, 
Phys. Rev. A {\bf 69}, 045601 (2004); 
{\it ibid} {\bf 70}, 013606 (2004); 
{\it ibid} {\bf 72}, 025602 (2005). 

\bibitem{11} M. Girardeau, J. Math. Phys. {\bf 1},
516 (1960); M. Girardeau, Phys. Rev. {\bf 139}, B500 (1965);
L. Tonks, Phys. Rev. {\bf 50}, 955 (1936); G. E. Astrakharchik, D. 
Blume, S. Giorgini, and 
B. E. Granger, J. Phys. B {\bf 37}, S205 (2004).

\bibitem{ref}Z. Idziaszek and T.  Calarco,
 Phys. Rev. A {\bf 71}, 050701(R) (2005).

\bibitem{ref1} M. Olshanii, Phys. Rev. Lett. Phys. Rev. Lett. {\bf 81},
938 (1998); D. S. Petrov, M. Holzmann, and G. V. Shlyapnikov, Phys. Rev.
Lett.  {\bf 84}, 2551 (2000); D. S. Petrov and G. V. Shlyapnikov, Phys.
Rev. A {\bf 64}, 012706 (2000); B. Tanatar, A. Minguzzi, P. Vignolo, and
M. P.  Tosi, Phys. Lett. A {\bf 302}, 131 (2002).


\bibitem{jz} D. M. Jezek, M. Barranco, M. Guilleumas, 
R. Mayol, and M. Pi, Phys. Rev. A {\bf 70},
043630 (2004).

\bibitem{pi} M. Pi, X. Vi\~nas, F. Garcias, and M. Barranco, 
Phys. Lett. B {\bf 215}, 5 (1988). 

\bibitem{sala-boris} L. Salasnich and B. A. Malomed, 
Phys. Rev. A {\bf 74}, 053610 (2006). 

\bibitem{sk1} S. K.  Adhikari  and P. Muruganandam, J. Phys. B  
{\bf 35}, 2831 (2002); 
P. Muruganandam  and S. K.  Adhikari,  {\it ibid.} {\bf 36}, 2501 (2003). 
 
\bibitem{sk3}
S. K. Adhikari, Phys. Rev. A {\bf 65}, 033616 (2002);
S. K. Adhikari, Phys. Rev. E {\bf 65}, 016703 (2002);
F. Dalfovo and S. Stringari,  Phys. Rev. A  {\bf 53}, 2477 (1996);
F. Dalfovo and M. Modugno,  {\it ibid.} {\bf 61}, 023605 (2000).


\bibitem{bbcol}S. K. Adhikari,  Phys. Rev. A {\bf 63}, 043611 (2001).

\bibitem{sala-ideal} L. Salasnich, J. Math. Phys. {\bf 41}, 8016 (2000).  


\end{thebibliography}
\end{document}